\documentclass[aps,prl,reprint,superscriptaddress,flushbottom,floatfix,nobibnotes,nofootinbib,a4paper,showpacs,amsmath]{revtex4-1}
\usepackage{amssymb}
\usepackage{graphicx}
\newcommand{\be}{\begin{equation}}
\newcommand{\ee}{\end{equation}}
\newcommand{\bea}{\begin{eqnarray}}
\newcommand{\eea}{\end{eqnarray}}

\newcommand{\MS}{\overline{\rm MS}}

\newcommand{\nn}{\nonumber}

\def\lQ{\Lambda_{\mathrm{QCD}}}

\def\al{\alpha}

\def\siml{{\ \lower-1.2pt\vbox{\hbox{\rlap{$<$}\lower6pt\vbox{\hbox{$\sim$}}}}\ }}

\def\siml{{\ \lower-1.2pt\vbox{\hbox{\rlap{$<$}\lower6pt\vbox{\hbox{$\sim$}}}}\ }}
\def\bfnabla{\mbox{\boldmath $\nabla$}}
\def\bfsigma{\mbox{\boldmath $\sigma$}}

\begin{document}
\title
{Model independent determination of the muonic hydrogen Lamb shift and proton radius} 
\author{Clara Peset and Antonio Pineda}
\affiliation{Grup de F\'{\i}sica Te\`orica, Universitat
Aut\`onoma de Barcelona, E-08193 Bellaterra, Barcelona, Spain}
\date{\today}
\begin{abstract}
We obtain a model independent expression for the muonic hydrogen Lamb shift. This expression 
includes the leading logarithmic ${\cal O}(m_{\mu}\alpha^6)$ terms, as well as the leading ${\cal O}(m_{\mu}\alpha^5
\frac{m_{\mu}^2}{m_{\rho}^2})$ hadronic effects. 
The latter are controlled by the chiral theory, which allows for their model independent determination. In this paper we give the missing piece for their complete expression including the pion and Delta particles. Out of this analysis and the experimental measurement of the muonic hydrogen Lamb shift we determine the electromagnetic 
proton radius: $r_p=0.8412(15)$ fm. This number is at 6.8$\sigma$ variance with respect to the CODATA value.
The accuracy of our result is limited by uncomputed terms of 
${\cal O}(m_{\mu}\alpha^5\frac{m_{\mu}^3}{m_{\rho}^3},m_{\mu}\alpha^6)$. This parametric control of the 
uncertainties allows us to obtain a model independent estimate of the error, which is dominated by hadronic effects. 
\end{abstract}
\maketitle

The recent measurement \cite{Pohl:2010zza,Antognini:1900ns} of the muonic hydrogen Lamb shift, $E(2P_{3/2})-E(2S_{1/2})$,
\begin{eqnarray}
\label{DeltaEexp}
&&\Delta E^{\rm exp}=202.3706(23)\,\mathrm{meV}
\end{eqnarray}
and the associated determination of the electromagnetic proton radius:
$r_p= 0.84087(39)$ fm has led to a lot of controversy. The reason is that this number is 7.1$\sigma$ away from the CODATA value, $r_p= 0.8775(51)$ fm \cite{Mohr:2012tt}. 

In order to asses the significance of this discrepancy it is of fundamental importance to perform the computation (in particular of the errors)
in a model independent way. In this Letter we revisit the theoretical derivation of the muonic hydrogen Lamb shift with this aim in mind. In this respect, the use of effective field theories 
is specially useful. They help organizing the computation by providing with power counting rules that asses the importance of the different contributions. This becomes increasingly necessary as higher order effects are included. Even more important, these power counting rules allow to parametrically control the size of the uncalculated terms and, thus, give an educated estimate of the error. 
This discussion specially applies to the muonic hydrogen, as its
dynamics is characterized by several scales:
\be
\nn
m_{p} \sim m_{\rho},
\quad
m_{\mu} \sim m_{\pi} \sim m_r\equiv\frac{m_{\mu}m_p}{m_p+m_{\mu}},
\quad
m_r\al \sim m_e.
\ee
By considering ratios between them, the main expansion parameters are obtained:
\bea
&&
\label{ratio1}
\frac{m_{\pi}}{m_p} \sim \frac{m_{\mu}}{m_p} \approx \frac{1}{9}
\,,
\;
\frac{m_e}{m_r} \sim \frac{m_r\al}{m_r}\sim \frac{m_r\al^2}{m_r\al}\sim \al \approx \frac{1}{137}
\,.
\eea

This approach to the problem has been followed in \cite{Pineda:2002as,Pineda:2004mx,Nevado:2007dd} (see \cite{Pineda:2011xp} for a review of these computations) with a combined use of 
Heavy Baryon Effective Theory (HBET) \cite{Jenkins:1990jv}, Non-Relativistic QED (NRQED) \cite{Caswell:1985ui} and, specially, potential NRQED (pNRQED) \cite{Pineda:1997bj,Pineda:1997ie,Pineda:1998kn}. Particularly relevant for us is Ref. 
\cite{Pineda:2004mx}, which contains detailed information on the application of pNRQED to the muonic hydrogen. 
We refer to it for details (but an even more detailed account with extra results is in preparation \cite{Peset2}).

Since pNRQED describes degrees of freedom with $E \sim m_{r}\al^2$, 
any other degree of freedom with larger energy is integrated out. This implies treating the proton and muon in a non-relativistic fashion and 
integrating out the pion and Delta particles. This is achieved by matching HBET to NRQED. By integrating out the scale $m_{\mu}\al$, pNRQED is obtained 
and the potentials appear. Schematically the path followed is the following ($\Delta\equiv m_\Delta-m_p$): 
$$
{\rm HBET} \; \stackrel{(m_{\pi,\mu},\Delta)}{\Longrightarrow} {\rm NRQED} \; \stackrel{(m_{\mu}\al)}{\Longrightarrow} \; {\rm pNRQED}\,,
$$
and the resulting pNRQED Lagrangian reads
\bea
\label{lpnrqed}
&&L_{\rm pNRQED} =
\int d^3{\bf r} d^3{\bf R} dt S^{\dagger}({\bf r}, {\bf R}, t)
                \Biggl\{
i\partial_0 - { {\bf p}^2 \over2 m_r} 
\\
&&
\nonumber
- V ({\bf r}, {\bf p}, {\bfsigma}_1,{\bfsigma}_2) + e {\bf r} \cdot {\bf E} ({\bf R},t)
\Biggr\}
S ({\bf r}, {\bf R}, t)- \int d^3{\bf R} {1\over 4}  F_{\mu \nu}F^{\mu \nu}
\,,
\eea
where $S$ is the field representing the muonic hydrogen, ${\bf R}$ the center of mass coordinate and ${\bf r}$ the relative distance.
$V$ stands for the potential and admits an expansion in powers of $1/m_{\mu}$:
\be
V ({\bf r}, {\bf p}, {\bfsigma}_1,{\bfsigma}_2)
=
V^{(0)}(r)+{V^{(1)}(r) \over m_{\mu}}+{V^{(2)}(r) \over m_{\mu}^2}+\cdots\,.
\end{equation}
The potentials $V^{(i)}$ are obtained as an expansion in powers of $\al$ (as well as in powers of the other small ratios appearing in (\ref{ratio1})). 
They are obtained through matching to the underlying theory and can be found in \cite{Pineda:2004mx}. The spectrum is then obtained by 
the combined use of NR quantum mechanics perturbation theory and perturbative quantum field theory computations (if ultrasoft photons 
show up).  
As we have definite counting rules to asses the relative importance of the different terms we know when we can truncate the computation. 
The application of this program to the muonic hydrogen produces the contributions we use in our analysis, listed in Table \ref{table}. Most of the results were already available in the literature, we have reevaluated many and computed the missing term to the polarizability due to the Delta to obtain the accuracy we aim at in this paper. We now briefly discuss them focusing on the novel aspects. 

\begin{table}[htb]
\addtolength{\arraycolsep}{0.15cm}
$$
\begin{array}{|l||c|r l|}
 \hline 
{\cal O} (m_r \alpha^3)& V_{\rm VP}^{(0)} & 205.&\hspace{-0.5cm}00745  
\\ \hline
{\cal O} (m_r \alpha^4)& V_{\rm VP}^{(0)} & 1.&\hspace{-0.5cm}50795  
\\ \hline
{\cal O} (m_r \alpha^4)& V_{\rm VP}^{(0)} & 0.&\hspace{-0.5cm}15090 
\\ \hline
{\cal O} (m_r \alpha^5)& V_{\rm VP}^{(0)} & 0.&\hspace{-0.5cm}00752
\\ \hline
{\cal O} (m_r \alpha^5)& V^{(0)}_{LbL} &    -0.&\hspace{-0.5cm}00089
\\ \hline
{\cal O} (m_r \alpha^4\times \frac{m^2_\mu}{m^2_p})& V^{(2)}+V^{(3)} &   0.&\hspace{-0.5cm}05747
\\ \hline
{\cal O} (m_r \alpha^5)& V^{(2)}_{\rm soft}/{\rm ultrasoft} &   -0.&\hspace{-0.5cm}71903
\\ \hline
{\cal O} (m_r \alpha^5)& V^{(2)}_{\rm VP}        &  0.&\hspace{-0.5cm}01876
\\ \hline
{\cal O} (m_{\mu}\alpha^6\times \ln (\frac{m_{\mu}}{m_{e}}))& V^{(2)}; c^{(\mu)}_D &  -0.&\hspace{-0.5cm}00127
\\ \hline 
{\cal O} (m_{\mu}\alpha^6\times \ln \al)& V^{(2)}_{\rm VP}; c^{(\mu)}_D&  -0.&\hspace{-0.5cm}00454
\\[0.05cm] \hline \hline 
{\cal O} (m_r \alpha^4\times m^2_r r^2_p)& V^{(2)}; c^{(p)}_D; r^2_p & -5.&\hspace{-0.5cm}1975
\frac{r_p^2}{\rm fm^2}
\\[0.05cm] \hline
{\cal O} (m_r \alpha^5\times m^2_r r^2_p)& V_{\rm VP}^{(2)}; c^{(p)}_D; r^2_p & -0.& \hspace{-0.5cm}0283
\frac{r_p^2}{\rm fm^2}  
\\[0.05cm] \hline
{\cal O} (m_r \alpha^6\ln\al\times m^2_r r^2_p)& V^{(2)}; c^{(p)}_D; r^2_p & -0.&\hspace{-0.5cm}0014
\frac{r_p^2}{\rm fm^2}  
\\[0.05cm] \hline
{\cal O} (m_r \alpha^5\times \frac{ m_r^2}{m_{\rho}^2})& V_{\rm VP_{\rm had}}^{(2)}; d^{\rm had}_2 &  0.&\hspace{-0.5cm}0111(2)
\\ \hline
{\cal O} (m_r \alpha^5\times \frac{m^2_r}{m^2_{\rho}}\frac{m_{\mu}}{m_{\pi}})& V^{(2)}; c^{\rm had}_3 &  0.&\hspace{-0.5cm}0344(125)
\\ \hline
\end{array}
$$
\caption{{\it The different contributions to the muonic hydrogen Lamb shift in} meV {\it units.}}
\label{table}
\end{table}

The first 4 entries in Table~\ref{table} are the contributions to the Lamb shift associated to the electron vacuum polarization (VP) corrections 
to the static potential $V^{(0)}$ (see Eq. (13) in Ref.~\cite{Pineda:2004mx}). Specially difficult is the 4th entry, as it corresponds to the three-loop static potential and to the third order computation in perturbation theory. It was computed in \cite{Kinoshita:1998jf} (see also \cite{Ivanov:2009aa} for a small correction).
 
 The 5th entry corresponds to the contribution associated to the light-by-light (LbL) corrections to the static potential $V^{(0)}$ 
(see the $\delta\alpha$ term in Eq. (15) in Ref.~\cite{Pineda:2004mx}). It was obtained in \cite{Karshenboim:2010cq}, where a very long explanation was made to argue that the LbL 
contributions could be computed in the static approximation. This is evident in the effective field theory, as they correspond to a correction to the static potential, as already stated in Ref.~\cite{Pineda:2004mx}.
 
The 6th entry corresponds to the leading contribution due to the $\al/m^2$ and $1/m^3$ potentials. 
Even though it is formally ${\cal O}(m_r\al^4)$
it suffers an extra $m_{\mu}^2/m_p^2$ suppression. This explains why it is smaller than its naive natural size. 

The 7th entry is the sum of the ultrasoft correction (see Eq. (3.7) in Ref.~\cite{Pineda:1998kn} rescaling $m/2 \rightarrow m_r$) 
and the (one loop) $\al^2/m^2$ potential (see Eq. (B2) in Ref.~\cite{Pineda:1998kn}). 
This sum can be considered in an isolated way, as it produces a well defined contribution for the case of the muonium ($\mu e$), 
where there is no contribution due to the electron VP.

The 8th entry is the sum of the correction produced by 2nd order NR quantum mechanics perturbation theory of the $\al^2/r$ potential due to the electron VP 
together with the $\al/m^2$ and $1/m^3$ potentials, and the correction due to the $\al^2/m^2$ potential due to the electron VP. Again this sum constitutes a well defined set, as it 
can be parametrically distinguished from other contributions (formally through the number of light fermions). This contribution was first computed in \cite{pachucki1} 
and later corrected in \cite{Veitia,Borie:2012zz}. Nevertheless, a different number has been obtained in two recent analyses 
\cite{Jentschura:2011nx,KIK}. We confirm this last number, which is the one we quote in Table~\ref{table}.


These 8 entries give the complete ${\cal O}(m_r\al^5)$ result for a point-like proton. 
In this result we have kept the exact mass dependence.
The ${\cal O}(m_{\mu}\alpha^6)$ contribution is dominated by the logarithmic enhanced terms. 
Here, we compute the leading ones. We assign a general counting of $ m_r/m_p \siml \ln \al \sim \ln (m_e/m_{\mu})$.  Therefore, we only compute those 
contributions at leading order in the $m_r/m_p$ expansion, i.e. those where the proton is infinitely massive. In this approximation 
all the logs are generated by the electron VP (as the case without the electron would correspond to the standard hydrogen situation) producing the 9th and 10th entries of Table~\ref{table}, which we now discuss.

The 9th entry is due to the logarithmic enhanced ${\cal O}(\al^2)$ corrections to the $c_D^{(\mu)}$ Wilson coefficient (see Eqs.~(B.2/3) in Ref.~\cite{Pineda:1998kn}), 
which with this accuracy reads \cite{Barbieri:1973kk,Barbieri:1972as}
 (we introduce the finite term for completeness although we do not use it in our computations)
\bea
\nn
&&
c_{D,\MS}^{(\mu)}(\nu)=1+\frac{4\alpha}{3\pi}\ln\left(\frac{m_\mu^2}{\nu^2}\right)
+\left(\frac{\alpha}{\pi}\right)^2
\left(
\frac{8}{9} \ln^2\left(\frac{m_\mu}{m_e}\right)
\right.
\\
\nn
&&
-\frac{40}{27} \ln\left(\frac{m_\mu}{m_e}\right)
-\frac{1183}{324}+\frac{\pi ^2 }{6}\left(\frac{-32}{9}+18 \ln(2)\right)
-\frac{9 }{2}\zeta(3)
\\
&&
\left.+{\cal O}\left(\frac{m_e}{m_{\mu}}\right)
\right)
\,.
\label{cDmu}
\eea
It produces an $\al^3/m^2\times$log-potential, the expectation value of which gives the 9th entry.

The 10th entry is generated in the same way as the 8th entry but multiplied by 
the (logarithmic enhanced) ${\cal O}(\al)$ term of $c_D^{(\mu)}(\nu)$. The $\nu$ dependence gets regulated by the 
ultrasoft scale, which we set to $\nu=m_{\mu}\al^2$, producing the number we quote in Table \ref{table}. 

Both computations were considered before in Ref. \cite{pachucki1}. We agree with them for the significant digits given in this reference. It is also interesting to see that both contributions can be understood from a renormalization group analysis in some appropriate limit \cite{Pineda:2002bv}. This analysis also gives us 
information on the log structure of the recoil, $m_r/m_p$, corrections. At this order extra log-terms appear. 
Nevertheless, they are at most linear: ${\cal O}(m_{\mu}\alpha^6\frac{m_{\mu}}{m_p}\ln\al)$, i.e. there are no ${\cal O}(m_{\mu}\alpha^6\frac{m_{\mu}}{m_p}\ln^2\al)$ terms, contrary to the claim in Ref. \cite{Jentschura:2011nx}. 

For a point-like proton this computation would finish our analysis. The error would be due to uncomputed contributions of  ${\cal O}(m_{\mu}\al^6)$ and
${\cal O}(m_{\mu}\alpha^6\frac{m_{\mu}}{m_p}\ln\al)$. In Refs.~\cite{Jentschura:2011ck,Korzinin:2013uia} 
several terms of this order were computed. We use these analyses to estimate the error. Specially useful to us are 
the (a) and (d) entries in Table IV of the last reference. They are related with the large-log contributions discussed above but also include some finite pieces. We take the difference with the pure log-terms for the generic ${\cal O}(m_{\mu}\al^6)$ error. 1/2 of 
the sum of the 9th and 10th entries yields a similar error: $\sim 3$ $\mu$eV.

Since the proton is not point-like, we have to incorporate the finite-size effects 
due to its hadronic structure. These are encoded in the 
Wilson coefficients $c_D^{(p)}$, $d_2$ and $c_3$ of the NRQED Lagrangian,  
\be
\delta {\cal L}
=\frac{d_2}{m_p^2}F_{\mu\nu}D^2F^{\mu\nu}
-e\frac{c_D^{(p)}}{8m_p^2}N_p^{\dagger}\bfnabla \cdot {\bf E}N_p
+
\frac{c_3}{m_p^2}N^{\dagger}_pN_p\mu^{\dagger}\mu
\,,
\end{equation}
in the following way ($d_s$ can be found in Eq. (B.4) of \cite{Pineda:1998kn}):
\bea
c_{D,\MS}^{(p)}(\nu)
&\equiv& 1+\frac{4}{3}\frac{\alpha}{\pi}\ln\left(\frac{m_p^2}{\nu^2}\right)
+\frac{4}{3}r_p^2m_p^2+{\cal O}(\al^2)
\,,
\\
c_{3}(\nu)&\equiv&-\frac{m_p}{m_{\mu}}d_s(\nu)+c_{3}^{\rm had}+{\cal O}(\al^3)
\,,
\\
d_2&=&\frac{\al}{60\pi}+d^{\rm had}_2+{\cal O}(\al^2)
\,.
\eea
$\frac{4}{3}r_p^2m_p^2$, $c^{\rm had}_3$ and $d^{\rm had}_2$ are defined as the left-over 
Wilson coefficients after subtraction of the proton (pure-QED) point-like contributions.
 
All these hadronic corrections add to the delta potential and energy shift in a specific combination: 
\bea
D_d^{\rm had}&\equiv&-c^{\rm had}_3-16\pi\al d^{\rm had}_2+ \frac{2\pi\al}{3}r_p^2m_p^2
 \,,
\\
\nn
\delta V_{\rm had}^{(2)}(r) &\equiv&  \frac{1}{m_p^2}D_d^{\rm had}\delta^3({\bf r})
 \rightarrow  \Delta E = -\frac{D_d^{\rm had}}{m_p^2}\frac{1}{\pi}(\frac{m_r\al}{2})^3
\,.
\eea
This equation gives the leading hadronic correction to the energy shift, which is due to $r_p$ and listed in the 11th entry of Table \ref{table}. It is of ${\cal O} (m_r \alpha^4\times m_r^2r_p^2)$ with $r_p \sim \frac{1}{m_{\rho}^2}\ln m_{\pi}$. We also need the corrections proportional to $r_p$ to the next power in $\al$. They are due to the electron VP corrections to $\delta V_{\rm had}^{(2)}$, 
and to the 2nd order NR quantum mechanics perturbation theory of the $\al^2/r$ potential due to the electron VP together with $\delta V_{\rm had}^{(2)}$, similarly as for the 8th entry of Table \ref{table}. The result is listed in the 12th entry of Table \ref{table}. The next correction is of 
${\cal O} (m_r \alpha^6\ln\al\times m_r^2r_p^2)$. It has been computed in \cite{Friar:1978wv} and listed in the 13th entry of Table \ref{table}. We use 1/2 of this result for the error of the $r_p^2$ coefficient.

$d_2^{\rm had}$ encodes the hadronic vacuum polarization effects. They can be accurately determined from dispersion relations \cite
{Jegerlehner:1996ab} with small errors for our purposes. The contribution is in the 14th entry of Table \ref{table} (note that the proton point-like contribution is subtracted).

The energy shift proportional to $c_3^{\rm had}$ is usually named $\Delta E^{{\rm TPE}}$, the two-photon exchange contribution. 
When matching HBET to NRQED we integrate out the pion, but also the Delta, 
not only because the Delta is the closest resonance to the proton (see \cite{Hemmert:1996xg}), but also because in the large $N_c$ limit the Delta and proton become degenerate \cite{Dashen:1993ac}. Since $c_3^{\rm had}$ depends linearly on the muon mass, it is dominated by the infrared dynamics and diverges in the chiral limit. This produces an extra
$m_{\mu}/m_{\pi}$ suppression with respect to its natural size.
Therefore, the pure-chiral and Delta-related computation gives the leading and next-to-leading order effect, respectively:
\be
c^{\rm had}_{3,\rm LO} \sim \alpha^2 {m_{\mu} \over m_{\pi}}
\,,
\quad
c^{\rm had}_{3,\rm NLO} \sim \alpha^2 {m_{\mu} \over m_{\pi}}\times{m_{\pi} \over \Delta}
\,.
\ee
In order to estimate the size of the different contributions (and the associated error) we proceed as follows. 
We count $m_{\pi} \sim \sqrt{\lQ m_q}$ and $\Delta \sim \frac{\lQ}{N_c}$. We then have the double expansion 
$\frac{m_{\pi}}{\lQ} \sim \sqrt{\frac{m_q}{\lQ}}$ and $\frac{\Delta}{\lQ} \sim \frac{1}{N_c}$. We still have to determine the 
relative size between $m_{\pi}$ and $\Delta$. We observe that  $m_{\pi}/\Delta \sim N_c \sqrt{\frac{m_q}{\lQ}} \sim 1/2$.
Therefore, we associate a 50\% uncertainty to the pure chiral result. 
The Delta-related effects are large and constitute the leading corrections to the chiral limit. We compute them in this paper (actually, we have computed the $(m_qN_c/\Lambda)^n$ dependence 
to any order in $n$ and incorporated it in the result).
These corrections
are free of counterterms, yielding a pure prediction, and give (together with the strict chiral result) the nonanalytic behavior in the light quark mass $m_q$ and $1/N_c$ of $c^{\rm had}_3$.  New counterterms scale as $ \sim \al^2 \frac{m_{\mu}}{\lQ}$, which set the precision of our result:
\be
c^{\rm had}_{3} \sim \al^2 \frac{m_{\mu}}{m_{\pi}}\left[1+\#\frac{m_{\pi}}{\Delta}+\cdots\right]
+{\cal O}\left(\al^2 \frac{m_{\mu}}{\lQ}\right).
\ee

Once the Delta is incorporated in the computation, 
the splitting with the 
next resonances suggests a mass gap of order $\lQ \sim$ 500-770 MeV, depending whether one considers the Ropper resonance or the 
$\rho$. Therefore, we assign $\frac{m_{\pi}}{\lQ} \sim 1/3$ 
and $\frac{\Delta}{\lQ} \sim 1/2$, as the uncertainties of the pure chiral and the Delta-related contribution respectively. 
We add these errors linearly for the final error. A more quantitative 
estimate of the uncertainties would require the knowledge of more orders of the perturbative expansion to see the convergence pattern.

It is customary to split $c^{\rm had}_3$ into the Born (or Zemach, or $\langle r^3 \rangle$) and polarizability terms, 
and so we do: $c^{\rm had}_3=c_3^{\rm Born}+c_3^{\rm pol}$. The chiral/Delta correction to $c_3^{\rm Born}$ has been computed in  
Refs.~\cite{Pineda:2004mx,Peset2} producing the following energy shift:
\be
\label{Born}
\Delta E^{{\rm Born}}_{\rm LO+NLO}=10.08-1.81=8.3(4.3) \, {\rm \mu eV}\,.
\ee
The 1st term is the pure chiral correction. The 2nd term is the Delta-related contribution and corrects the result in 
Ref.~\cite{Pineda:2004mx}. We observe a rather good convergence. On the other hand our result is quite different with respect to standard values obtained from dispersion relations~\cite{Pachucki:1999zza,Carlson:2011zd}. One may wonder whether such difference is due to relativistic corrections. An estimate of the relativistic effects can be obtained from the analysis made in Ref.~\cite{pachucki1}, which, however,
is based on dipole form factor parameterizations. The difference between the relativistic and nonrelativistic expression was found to be small ($\sim \; 3 \mu$eV). It should be further investigated if this feature holds with different parameterizations. We relegate a broader discussion on this issue to Ref.~\cite{Peset2}. In the mean time we will stick to our model independent prediction from the effective theory.

\begin{figure}
\includegraphics[width=.45\textwidth,clip=]{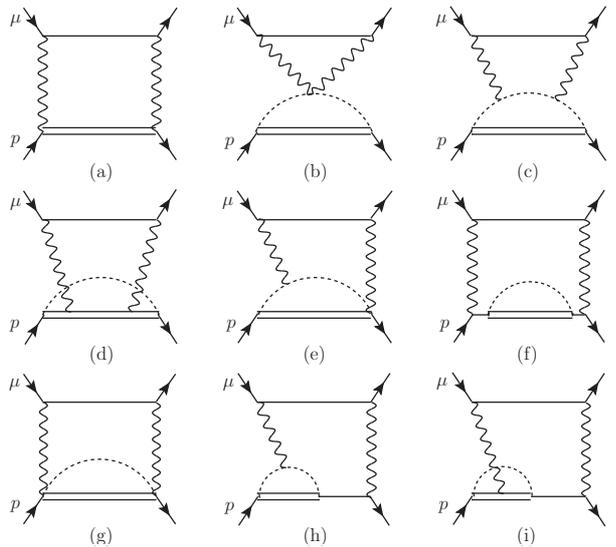}
\caption{\it Diagrams (up to symmetric permutations) involving the Delta particle needed to obtain the polarizability.
\label{fig:Delta}}
\end{figure}

The chiral correction to $c_3^{\rm pol}$ has been computed in \cite{Nevado:2007dd}. We have checked this result. From the power counting point of view, Delta effects are the most important corrections. 
Therefore, we compute the Delta-related contribution to the polarizability correction to the Lamb shift. This eliminates the major source of uncertainty of the polarizability contribution. In Fig.~\ref{fig:Delta} we show the diagrams that contribute to the polarizability due to the 
Delta particle. Overall, we obtain the following energy shift from the polarizability effects 
\begin{equation}
\Delta E^{\rm pol}_{\rm LO+NLO}=
18.51+7.67=26.2(10.0) \; {\rm \mu eV}
\,.
\label{pol}
\end{equation}
The first term is the pure chiral correction, already obtained in \cite{Nevado:2007dd}. The second term is the Delta-related contribution 
and it is new. It is smaller than the pure chiral result (as in the Born case), which we find reassuring. We have numerically checked that the 
$m_{\mu} \rightarrow 0$ limit coincides with Eq. (51) of \cite{Pineda:2004mx}. Corrections to this result are parametrically suppressed by 
a factor $m_{\mu}/\lQ$.  There also exists a computation, using a relativistic version of chiral perturbation theory, for the chiral-related term \cite{Alarcon:2013cba}.  Such computation treats the baryon relativistically. This may 
jeopardize the power counting by introducing, in the same footing, some subleading contributions. It is usually said that such subleading effects may give an estimate of higher order effects in HBET. Nevertheless, such computations also assume that a theory with only baryons and pions is appropriate at the proton mass scale (actually the very fact that those are the right degrees of freedom at these scales could be debatable). This should be taken with due caution. 
Still, it would be desirable to have a deeper theoretical understanding of this difference, which may signal that relativistic corrections are 
important for the polarizability correction. In any case, their result differs from our chiral result by 
around 50\%. This is around 1.5 times the error we use for the chiral contribution once the Delta is incorporated in the calculation, which we 
consider reasonable. 

Combining the Born and polarizability contributions we obtain the two-photon exchange term (listed in the 15th entry in Table \ref{table}):
\be
\Delta E^{\rm TPE}_{\rm LO+NLO}=
28.59+5.86=34.4(12.5) \; {\rm \mu eV}
\,.
\label{TPE}
\end{equation}
We would like to emphasize that this result is a pure prediction of the effective theory. It is also the most precise expression that can be obtained in a model independent way, since ${\cal O}(m_{\mu}\alpha^5\frac{m_{\mu}^3}{\lQ^3})$ effects are not controlled by the chiral theory and would require new counterterms. This problem is not (nor it can be) solved by the analysis of \cite{Birse:2012eb}, where the low energy behavior of the forward virtual Compton tensor was computed to ${\cal O}(p^4)$, since a model dependent form factor was used, not only at the $\rho$-mass scale, but also at the chiral scale. Finally, unlike in the Born case, dispersion relation analyses require subtractions. As it has been emphasized in Ref.~\cite{Hill:2011wy}, this introduces dependences on unmeasured amplitudes, which necessarily require modelling, making both the absolute value and the error analysis of these determinations model dependent, and, thus, arbitrary to a large extent. This leaves our analysis as the only one that eliminates all model dependence.  

Summarizing all contributions, our final prediction for the Lamb shift reads
\bea
\label{El1}
&&
\Delta E^{\rm this\,work}=206.070(13)-
5.2271(7)
\frac{r_p^2}{\mathrm{fm^2}}\,\mathrm{meV}
\\
&&=
206.0243(30)
-5.2271(7)
\frac{r_p^2}{\mathrm{fm^2}}+0.0455(125)\,\mathrm{meV}
\,.
\eea
In the last equality the first term is the pure QED result, and its error is the estimate of the ${\cal O}(m_{\mu}\al^6)$ effects. The error of the coefficient of the term proportional to $r_p^2$ is the 
estimated size of the ${\cal O}(m_{\mu}\al^6 (m_{\mu}r_p)^2)$ terms. The last term encodes the $r_p$-independent hadronic effects. The error is the assigned uncertainty due to unknown terms of ${\cal O}(m_{\mu}\alpha^5\frac{m_{\mu}^3}{m_{\rho}^3})$. 
Using 
Eq.~(\ref{DeltaEexp}) we obtain  
\begin{equation}
r_p=0.8412(15) \, \mathrm{fm},
\end{equation}
where the theoretical and experimental errors have been combined in quadrature. Nevertheless, the latter is completely subdominant with respect to the total error, which is fully dominated by the hadronic effects. 

Our central value is basically equal to the one quoted in \cite{Antognini:1900ns} (even though some individual terms are quite different) 
but has significantly larger errors. The main reason is that the error associated to the two-photon exchange contribution is larger in our case, as it is the most one can do without model dependent assumptions. Nevertheless, we emphasize that the proton radius puzzle survives our model independent analysis, which yields a 6.8$\sigma$ discrepancy with respect to the CODATA value.

\begin{acknowledgments}
This work was supported by the Spanish 
grants FPA2010-16963 and FPA2011-25948, and the Catalan grant SGR2009-00894.
\end{acknowledgments}

\end{document}